\documentclass[aps,prb,amsmath,amssymb,footinbib,showpacs,twocolumn]{revtex4-2}
\usepackage{amsmath}
\usepackage{amssymb}
\usepackage{amsthm}
\usepackage{setspace}
\usepackage{graphicx}
\usepackage{braket}
\usepackage{mathrsfs}
\usepackage{float}
\usepackage[colorlinks = true,linkcolor = blue,urlcolor  = blue,citecolor = blue,anchorcolor = blue]{hyperref}
\usepackage[utf8]{inputenc}
\usepackage[english]{babel}
\usepackage{bm}

\begin{document}

\title{Extended Kohler's scaling, a low temperature anomaly and Isosbestic point in the charge density wave state of 1T-VSe$_2$}
\author{Sonika$^1$}
\author{Sunil Gangwar$^1$}
\author{Pankaj Kumar$^1$}
\author{A. Taraphder$^2$}
\author{C.S. Yadav$^1$}
 \email{shekhar@iitmandi.ac.in}

\affiliation{$^1$ School of Physical Sciences, Indian Institute of Technology Mandi, Kamand, Mandi-175075 (H.P.) India}

\affiliation{$^2$ Department of Physics, Indian Institute of Technology Kharagpur, West Bengal-721302, India}

\begin{abstract} 

1T-VSe$_2$ is a narrow band transition metal chalcogenide that shows charge density wave (CDW) state below $\textit{T}_{CDW}$ = 110 K. Here, we have explored the relevance of Kohler's rule and the thermal transport properties of VSe$_2$ across the CDW state. The magnetoresistance (MR) follows Kohler's rule above $\textit{T}_{CDW}$, while an extended Kohler's rule is employed below $\textit{T}_{CDW}$. Interestingly, we observed an anomaly in MR at \textit{T} = 20 K, below which MR value decreases on lowering temperature. This anomaly is also reflected in the slope ($\kappa$) of Kohler's plots and the relative change in the thermal excitation induced carrier density ($n_T$) also. The $\textit{T}_{CDW}$ remains largely unaffected in both electrical resistivity (\textit{$\rho(T)$}) and longitudinal Seebeck coefficient ($\it{S_{xx}}$) even under a strong magnetic field of 14 Tesla. However, the application of magnetic field enhances the peak intensity of $\it{S_{xx}}$ at \textit{T} $\sim$ 60 K. Additionally, $\it{S_{xx}(T)}$ curves measured at different fields exhibit a crossover at \textit{T} = 20 K, which suggest the existence of unique feature in the CDW state of VSe$_2$ \textit{i.e.} a locally exact isosbestic point. 

\end{abstract}

\maketitle

\section{Introduction}

The electronic transport in the charge density wave (CDW) systems is of significant interest from both theoretical and experimental point of views ~\cite{naito1982electrical, ghiringhelli2012long, achkar2016orbital, wang2021pressure, dong2022record}. The CDW state is characterized by the periodic modulations of conduction electron density in metals, and the Peierls distortion of the crystal lattice at low temperatures \cite{peierls1955quantum, RevModPhys.60.1129, gruner2018density}. The formation of the superlattice structure gives rise to a energy gap and the nesting of the Fermi surface (FS). Consequently, the FS undergoes modification in both shape and topology. These changes in FS, have a direct influence on the transport properties of the material. In some compounds, CDW gap opens across the entire FS, causing a metal-insulator transition \cite{gama1993interplay, pouget1983evidence}. However, in many other compounds such as 1T-TiSe$_2$ and NbSe$_3$, only a portion of the FS develops energy gap during CDW transition \cite{PhysRevB.14.4321, PhysRevB.16.3443}. The existence of CDW instabilities in group-V transition metal dichalcogenides (TMDs) is widely recognized. Among these TMDs, 1T-VSe$_2$ shows CDW below $\sim$110 K.  The VSe$_2$ has been studied for its CDW properties and intrinsic electronic structure. VSe$_2$ exhibits a partially filled 3d orbital which is extremely narrow compared to 4d and 5d bands, and hence enhanced electron-electron interactions \cite{zhou2016vanadium}. This difference in d-bands can greatly influence the physical properties of the material \cite{taraphder2011preformed, koley2014preformed}. Recently, several intriguing phenomenon were observed in the magnetotransport studies of rare earth trichalcogenide LaTe$_{3}$, manifesting the interplay of CDW and field induced modification of electronic structure \cite{PhysRevB.104.155147}. Rare earth trichalcogenide LaTe$_{3}$ show CDW transition above room temperature, with a very limited studies above its CDW transition. These studies motivated us to explore the role of magnetic field on the electronic transport on VSe$_2$ across the CDW transition. 
 
One of these properties is the magnetoresistance (MR), which is defined as the change in electrical resistance induced by applied magnetic field, offers valuable insights into the FS, carrier scatterings, and exotic phases of the system \cite{pippard1989magnetoresistance, PhysRevB.99.035142, abrikosov2017fundamentals}. Typically, in metals, MR follows a quadratic field dependence, as predicted by the Drude formula. However, the MR in topological systems like Dirac/Weyl semimetal \cite{sonika2021planar, PhysRevB.84.220401, PhysRevLett.106.217004}, show linear behavior with the magnetic field. Additionally, quasi-linear MR has been reported in non-magnetic silver chalcogenides \cite{xu1997large, PhysRevB.108.245141}. In 1938, Kohler showed that the MR is function of the ratio of the magnetic field \textit{H} to the zero-field resistivity $\rho_{xx}(0)$ \textit{i.e.} MR = f($\textit{H}/\rho_{xx}(0)$) \cite{kohler1938magnetischen}. Interestingly, this rule, although derived from the semiclassical Boltzmann theory, has been found to apply not only to metals but also to several strongly correlated systems like semimetals, heavy fermions, cuprate supeconductors as well \cite{PhysRevLett.125.176601, PhysRevB.97.245101, PhysRevLett.114.117201, PhysRevB.96.165145, PhysRevB.97.235132, PhysRevLett.113.177005}. However, deviations from Kohler's rule have also been there owing to various phenomena such as Fermi surface reconstruction, anisotropic effects, magnetic breakdown, and other carrier scattering mechanisms like phonon scattering or electron-electron interactions \cite{PhysRevLett.124.167602, PhysRevLett.115.166602, PhysRevB.92.060505, PhysRevLett.75.1391, PhysRevB.95.085202, PhysRevB.53.8733, PhysRevB.102.035164, PhysRevB.96.195107, PhysRevB.102.035164, harimohan2019magneto, PhysRevB.89.144512, li2022violation, li2005effect, xue2020thickness, kolincio2019charge, yasuzuka2005violation, PhysRevX.11.041029}. Specifically, Kohler's rule is expected to be violated in materials characterized by low carrier density, as thermal excitation leads to a significant change in carrier density \cite{kolincio2019charge, xue2020thickness, li2005effect, yasuzuka2005violation}. 

We have explored the electrical and thermal transport properties of VSe$_2$ across the CDW state in the presence of magnetic field. We observed that the MR follows Kohler's rule for $\textit{T} > \textit{T}_{CDW}$ and an extended version of Kohler's rule for $\textit{T} < \textit{T}_{CDW}$. Interestingly, the MR value increases with increasing temperature up to \textit{T} $\sim$ 20 K, followed by a typical decreasing trend at higher temperatures. The Kohler slope ($\kappa$) and the relative change in carrier density induced by thermal excitations ($n_T$) also exhibit abrupt changes at \textit{T} = 20 K. Furthermore, the longitudinal Seebeck coefficient ($\it{S_{xx}}$) curves measured at different magnetic fields, show a crossover at the same temperature pointing at the existence of locally exact isosbestic point. Additionally, the Hall resistivity ($\rho_{xy}$), transverse Seebeck coefficient ($\it{S_{xy}}$), and the Hall angle ($\theta_H$) also ascertain this anomaly at \textit{T} = 20 K.

\section{Experiments}

Single crystalline samples of VSe$_2$ were prepared using chemical vapor transport method. First, the polycrystalline samples were prepared by solid state reaction route, taking V turnings (99.7 $\%$) and Se shots (99.99 $\%$) in stoichiometric ratio with 5 $\%$ excess Se to minimize the self intercalation of V into interstitial sites. The polycrystalline VSe$_2$ was grounded, pelletized and sealed in evacuated quartz tube with pressure lower than ${10}^{-4}$ mbar with 0.08 g iodine as a transport agent. In the second step, the tube containing the sample was heated to ${750}^o$C for 15 days with a gradient of ${90}^o$C between hot and cold end of the tube. We could obtain a large number of single crystals of dimensions 2 mm x 5 mm. 
We performed room temperature XRD measurements on powdered sample using Rigaku Smartlab X-Ray Diffractometer. The analysis was carried out using Full Prof Rietveld refinement method which confirms the hexagonal structure (space group: {\it P}$\bar{3}$m1) for VSe$_2$. To investigate electronic transport properties, we used Quantum Design Physical Property Measurement System (PPMS) applying an alternating current of 1 mA with a frequency of 128.3 Hz. For the measurement of Seebeck coefficient, we used a home built experimental set up integrated with the PPMS \cite{sharma2020experimental}. 
 
\section{Results and Discussion}
\begin{figure*}
\includegraphics[width= 16 cm, height = 12 cm]{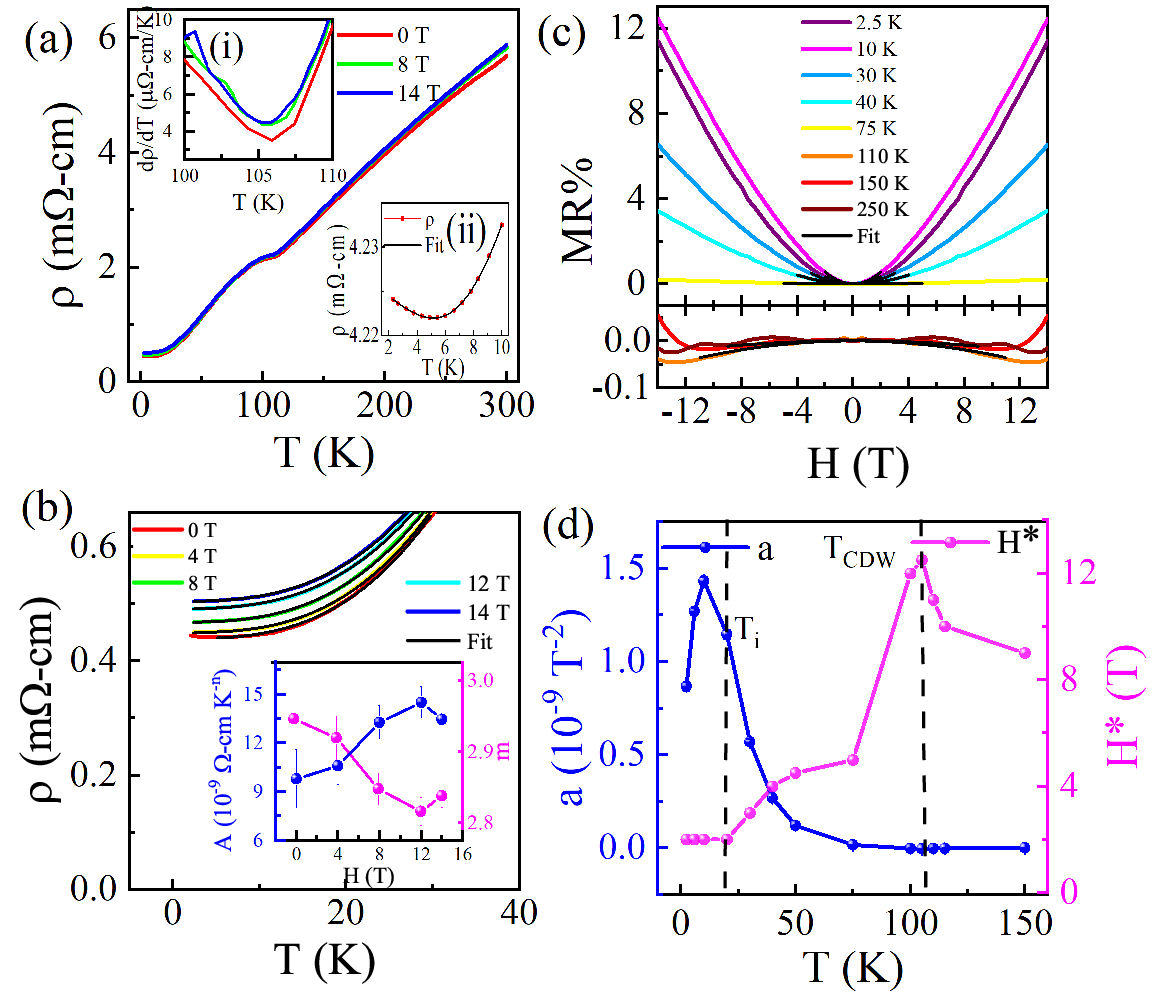}
\caption{(a) The \textit{$\rho (T)$} at various magnetic fields. Inset (i) shows the derivative of resistivity, inset (ii) depicts Kondo fit of \textit{$\rho (T)$} (b) Low temperature \textit{$\rho (T)$} fit the expression $\rho = \rho_0 + A\textit{T}^m$. Insets show the magnetic field dependence of (i) coefficient \textit{A}, and (ii) Power exponent \textit{m}. (c) MR measured for \textit{T} = 2.5 - 75 K (upper panel) and \textit{T} = 110 - 250 K (lower panel) when $\it{H} \perp \it{I}$. (d) Temperature dependence of parameters 'a' and \textit{H$^*$} extracted from parabolic fit of MR.}
\label{fig:Figure1}
\end{figure*}

The electrical resistivity (\textit{$\rho(T)$}) of 1T-VSe$_2$ is measured for temperature (\textit{T}) range 2 - 300 K under various magnetic fields (\textit{H}) as depicted in figure \ref{fig:Figure1} (a). The \textit{$\rho(T)$} exhibits a metallic behavior above the CDW transition ($\textit{T}_{CDW}$ $\sim{110}$ K). The \textit{d$\rho/dT$} (inset (i) of the figure \ref{fig:Figure1} (a)) shows that there is no appreciable change in $\textit{T}_{CDW}$ with applied magnetic field. Additionally, zero field \textit{$\rho(T)$} displays an upturn at low \textit{T} which get suppressed with application of magnetic field, indicative of a Kondo-like behavior (inset (ii) of the figure \ref{fig:Figure1} (a)) \cite{barua2017signatures}. Generally, the Kondo effect manifests in metals or alloys containing a magnetic impurity atom or ion embedded in a non-magnetic host lattice. At low \textit{T}, the magnetic moments of the impurity atom interact with the conduction electrons of the host that results in the scattering of conduction electrons by the impurity. We have fitted low \textit{T} (2 - 10 K), zero field \textit{$\rho (T)$} using the equation \cite{PhysRev.158.570, barua2017signatures};  

\begin{equation}
\rho (T) = \rho_b + pT^r + q \rho_H
\label{eq1}
\end{equation}

where first term on the right side of equation \ref{eq1} \textit{i.e.} $\rho_b$ is \textit{T} independent term, second term accounts for the contribution from electron-phonon interactions and the third term represents the Hamann expression given by,

\begin{equation}
\rho_H = \Biggl\{1-ln\left(\frac{T}{T_K}\right) \left[ln^2\left(\frac{T}{T_K}\right) + s (s+1) \pi^2 \right]^\frac{-1}{2}\Biggr\}
\label{eq2}
\end{equation}

Here, \textit{T$_K$} refers to the Kondo temperature and \textit{s} denotes the spin of magnetic impurity. The values \textit{p} and \textit{r} obtained from the fit are 1.37$\times$10$^{-6}$ and 1.53 respectively. The \textit{T$_K$} and spin \textit{s} values obtained from the fitting are $\sim{7.52}$ K and 0.12 respectively, which are comparable with the values in literature \cite{barua2017signatures}.  

Further, we have fitted \textit{$\rho (T)$} behavior at different \textit{H}, for \textit{T} = 2 - 30 K using the expression $\rho = \rho_0 + A\textit{T}^m$ which is shown in figure \ref{fig:Figure1}(b). The inset of this figure illustrates the \textit{H} dependence of the coefficient \textit{A} and the power exponent \textit{m}, showing the increase in \textit{A} and decreases in with the increase in value of \textit{H}. Although it is natural to expect a $\textit{T}^2$ dependence of \textit{$\rho (T)$} for Fermi liquid system at low temperature, the VSe$_2$ system exhibits a $\sim \textit{T}^3$ dependence for \textit{T} = 2 - 30 K. Earlier reports on 1T-VSe$_2$, have also shown non-Fermi liquid behavior at low temperature and with the varied values of the exponent \cite{toriumi1981galvanomagnetic, van1978magnetic, yadav2008transport}. Although, the $\textit{T}^3$ dependence of resistivity is observed in Arsenic, Antimony, Ag$_x$PdTe$_2$ and some Eu based valence fluctuating systems \cite{bansal1973temperature, PhysRevB.108.245141, patil1990anomalies}, we believe that such behavior in 1T-VSe$_2$ depends on the quality of the single crystal and scattering in the low temperature commensurate CDW state. However, the exact reason for this dependence is hitherto unknown to us. 

Figure \ref{fig:Figure1} (c) shows the magnetoresistance (MR\% = $(\rho(H) - \rho(0)) \times 100/\rho(0)$) for \textit{T} = 2.5 - 75 K (upper panel) and \textit{T} = 110 - 250 K (lower panel). We observed a positive MR for \textit{T} = 2.5 - 75 K. It should be emphasized that MR turns negative around $\textit{T}_{CDW}$ and again becomes positive for \textit{T} $>$ 150 K. Negative MR around $\textit{T}_{CDW}$ may be attributed to the changes induced by external \textit{H} in the electronic structure and FS of the material. The MR data reveals a quasi-linear dependence on \textit{H} for \textit{T} = 2.5 - 75 K. However, for specific ranges of \textit{H}, the MR data exhibits a parabolic dependence which varies for different \textit{T}. We fitted the MR data (shown by solid black lines) using the expression MR = a$\textit{H}^2$ for entire \textit{T} range. The coefficient 'a' and field \textit{H$^*$} up to which the MR data is parabolic reveal an interesting \textit{T} dependence of electronic transport (figure \ref{fig:Figure1} (d)). The \textit{H$^*$(T)} is constant up to  $\textit{T}_i$ $\sim$ 20 K, before rising to maximum at $\textit{T}_{CDW}$. Simultaneously, the coefficient 'a' also exhibits a peak below $\textit{T}_i$ $\sim$ 20 K, and eventually reaching to minimum at $\textit{T}_{CDW}$. 

\begin{figure*}
\includegraphics[width= 18 cm, height = 11 cm]{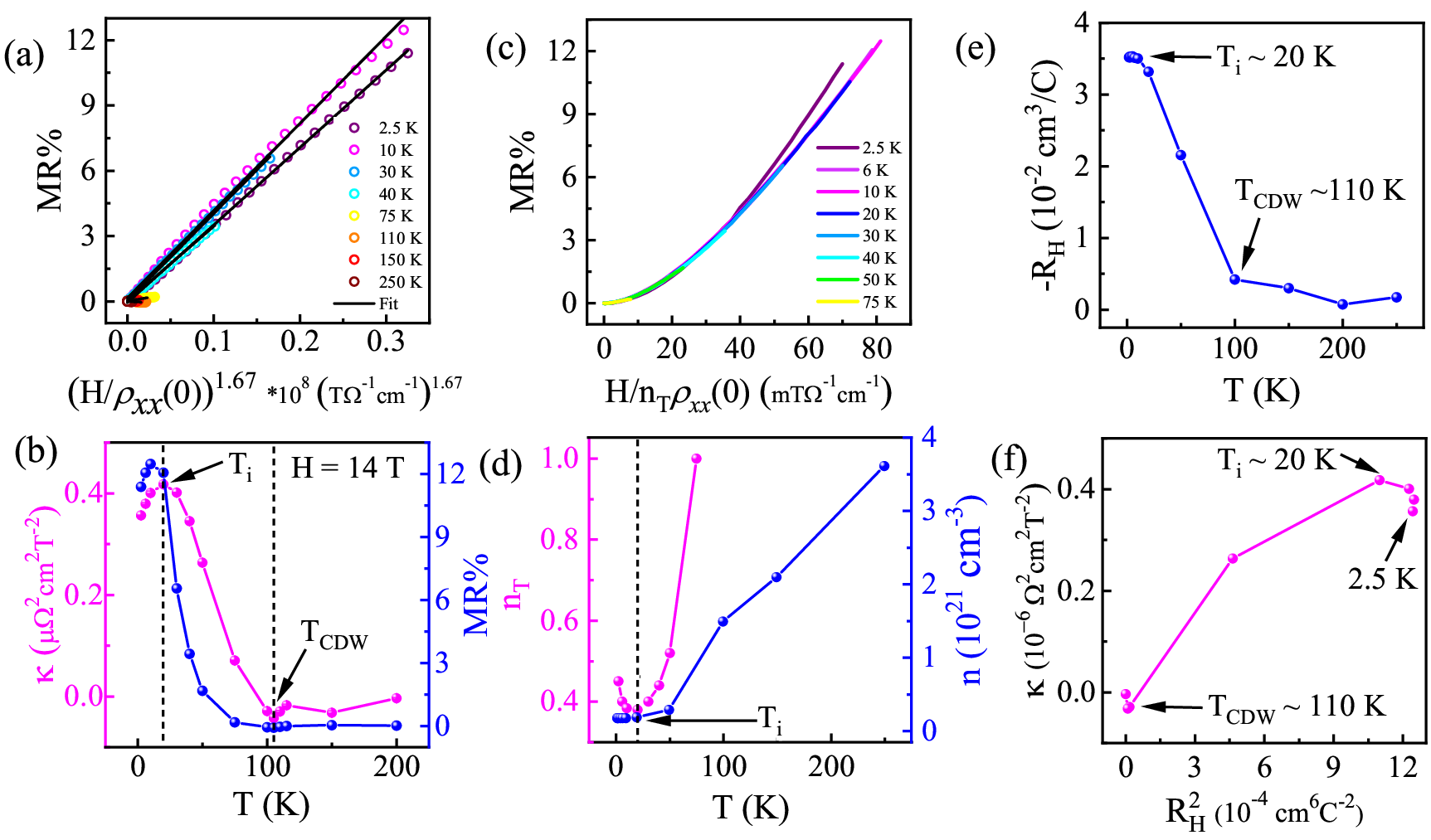}
\caption{Kohler analysis of MR in 1T-VSe$_2$ crystal (a) MR plotted against $(\textit{H}/\rho_{xx}(0)^{1.67}$. The black solid lines show the linear fit (b) Temperature dependence of Kohler slopes ($\kappa$) (left) and MR (\textit{H} = 14 Tesla) (right) (c) Extended Kohler's rule plots for MR curves plotted in (a) (d) Temperature dependence of $n_T$ (left) and carrier concentration n (right) (e) Temperature dependence of Hall coefficient $R_H$ (f) Kohler's slopes ($\kappa$) plotted against squared Hall coefficient ($R_H^2$).}
\label{fig:Figure2}
\end{figure*} 

Further, we analyzed the MR data by employing Kohler’s scaling at different temperatures. According to Kohler's rule, MR, which is the change in $\rho$ in an applied field, depends on $\omega_c \tau$, where $\omega_c (\varpropto \it{H}$) is the cyclotron frequency and $\tau(T) (\varpropto 1/\rho(T)$) is the relaxation time. Therefore, MR is a function of $\textit{H}/\rho_{xx}(0)$ \cite{ziman2001electrons}. We observed that the MR follows a power law behavior with exponent varying between 1.5 to 1.7 (Figure \ref{fig:Figure1}(c)). Therefore, we choose an aggregate exponent of $\sim{1.67}$ and plotted MR data against ($\textit{H}/\rho_{xx}(0))^{1.67}$) in figure \ref{fig:Figure2} (a) which shows nearly linear curves for all \textit{T}. The MR curves at different \textit{T} overlap onto a single curve above $\textit{T}_{CDW}$, suggesting a single scattering time above $\textit{T}_{CDW}$. However, for $\textit{T}$ $<$ $\textit{T}_{CDW}$ the Kohler's rule is violated which is expected, as metallic CDW in the dichalcogenides is associated with a lattice distortion and opening of a partial energy gap at the Fermi surface. We have plotted the \textit{T} dependence of Kohler's slopes ($\kappa$ = d(MR)/d($(\textit{H}/\rho_{xx}(0))^{1.67}$)), extracted from linear fits of MR curves and MR$\%$ values at \textit{H} = 14 Tesla in figure \ref{fig:Figure2} (b). Both MR and slopes $\kappa$ increase with the increase in \textit{T} until $\textit{T}_i$ $\sim$ 20 K and then decrease and attain a minimum at $\textit{T}_{CDW}$ and becomes almost constant above it. 

Recently, an extended Kohler’s rule \cite{xu2021extended} proposed that a temperature-dependent multiplier to MR or to $\textit{H}/\rho_{xx}(0)$ could cause different MR curves to overlap onto one single curve. Therefore, we implemented the extended Kohler's rule in CDW state, using MR = \textit{f}($\textit{H}/n_T\rho_{xx}(0)$), where \textit{n$_T$} depicts \textit{T} dependence of carrier density. For different MR curves to overlap onto a single curve, we multiplied them with different \textit{n$_T$} values and the plots for MR as a function of $\textit{H}/n_T\rho_{xx}(0)$ are shown in figure \ref{fig:Figure2} (c). The corresponding \textit{n$_T$} values at different \textit{T} below $\textit{T}_{CDW}$ are shown in figure \ref{fig:Figure2} (d) along with the carrier concentration obtained from Hall measurement. All the MR curves were normalized to MR curve at \textit{T} = 75 K and we observed that the value of \textit{n$_T$} decreases on lowering \textit{T} down to $\textit{T}_i$ $\sim$ 20 K and after that it rises with further decrease in \textit{T}. This behavior of \textit{n$_T$} reflects the relation with carrier density which is evident when we compare the carrier concentration obtained from Hall measurement. Since \textit{n$_T$} is associated with the Fermi level and electronic band structure, the transition at $\textit{T}_i$ $\sim$ 20 K could potentially be linked with Fermi surface and electronic band structure \cite{xu2021extended}. The temperature dependence of Hall coefficient R$_H$ of 1T-VSe$_2$ system is presented in figure \ref{fig:Figure2} (e) which shows a distinct increase in R$_H$ and hence decrease in carrier concentration below $\textit{T}_{CDW}$ and R$_H$ is almost constant below $\textit{T}_i$ $\sim$ 20 K.

In Figure \ref{fig:Figure2}(a), the MR is presented as MR = $\kappa$$(\textit{H}/\rho_{xx}(0))^{1.67}$. Additionally, the classical orbital MR can be represented as MR $\sim$ ($\mu\textit{H})^2$, where $\mu$ denotes the mobility of the carriers. Consequently, it was anticipated that $\kappa$ = $(\rho_{xx}(0)\mu)^2$ = $1/(ne)^{2}$ = $R_H^2$, with $R_H$ = -1/(ne). So, the extended Kohler’s rule can be rewritten as MR = \textit{f}($R_H\textit{H}/\rho_{xx}(0)$). Figure \ref{fig:Figure2}(f) exhibits the relationship between $\kappa$ and $R_H^2$, revealing two key features: (i) the non-monotonic relationship of $\kappa$ and $R_H^2$ and (ii) the $R_H$ value surpasses $\kappa$ by nearly 2-3 order of magnitude. Assuming the dominance of a single type of carriers in governing the transport properties of the 1T-VSe$_2$ crystals and accounting for the non-uniform scattering rate on the Fermi surface, the experimentally measured $R_H$ and $\kappa$ can be expressed as $R_H = \frac{1}{ne}\left(\frac{\langle\tau^2\rangle}{\langle\tau\rangle^2}\right)$ and $\kappa = \frac{1}{(ne)^2} \left(\frac{\langle\tau^3\rangle}{	\langle\tau\rangle^3}\right)$, respectively \cite{hurd2012hall, pippard1989magnetoresistance}. Here $\langle  \rangle$ represents an average over the Fermi surface, and $\tau$ denotes the carrier scattering time. If $\tau$ exhibits only minor variations across the Fermi surface, then $\kappa$ = $R_H^2$ = $(ne)^{-2}$. However, if $\tau$ exhibits considerable variations on the Fermi surface, we would expect $\kappa/R_H^2 = \langle\tau^3\rangle \langle\tau\rangle/ \langle\tau^2\rangle^2$. Evidently, the non-monotonic relationship between $\kappa$ and $R_H^2$ observed in figure \ref{fig:Figure2}(f) demonstrates that the electron scattering rate 1/$\tau$ significantly varies at different points on the Fermi surface. A distinct dip is observed in $\kappa-R_H^2$ curve at $\textit{T}_{CDW}$ which is associated with lattice distortion and reconstruction of electronic Fermi surface. Both of these effects lead to the abrupt change in anisotropy of electron scattering time, either directly or indirectly through electron-phonon interactions. Also, $R_H$ and $\kappa$ are associated with the electron scattering time anisotropy on the Fermi surface. Therefore, a sizeable change in $\kappa$ in the vicinity of $\textit{T}_{CDW}$ indicates that $\kappa$ is primarily influenced by anisotropy of electron scattering time. Additionally, \textit{T} dependence of MR (figure \ref{fig:Figure2}(b)), shows a minimum at $\textit{T}_{CDW}$ which arises because MR is closely related to the Kohler slope (MR $\varpropto$ $\kappa$).

The violation of Kohler's rule is not limited to charge density wave systems, but also observed in topological materials, high-$\textit{T}_C$ cuprates, iron-based superconductors, and heavy fermion intermetallics \cite{PhysRevB.59.14723, PhysRevLett.86.4652, PhysRevB.81.184519, nakajima2007non}. There are numerous ways to explain the deviation from Kohler's rule, out of which the most widely used is a modified Kohler's rule given by

\begin{equation}\\
MR = \frac{(\rho(H) - \rho(0))}{\rho(0)} = \gamma_H tan^2 \theta_H
\label{eq3}
\end{equation}

where $\theta_H$ = tan$^{-1}(\rho_{xy}/\rho_{xx})$ is the Hall angle and $\gamma_H$ determines the ratio of hole and electron mobility. This scaling method (equation \ref{eq3}) is valid for two-band compensated materials. An alternative scaling technique has been proposed by Xu $\textit{et al.}$ \cite{xu2021extended} tailored for single-band systems with low mobility, characterized by an anisotropic Fermi surface, non-compensated two-band and multi-band systems. For low mobility, $\rho_{xx} \approx \rho_{0}$, therefore equation \ref{eq3} can be rewritten as 

\begin{equation}\\
MR = \gamma_H (\rho_{xy}/\rho_{0})^2
\label{eq4}
\end{equation}

\begin{figure} [t]
\includegraphics[width = \columnwidth, height = 16 cm]{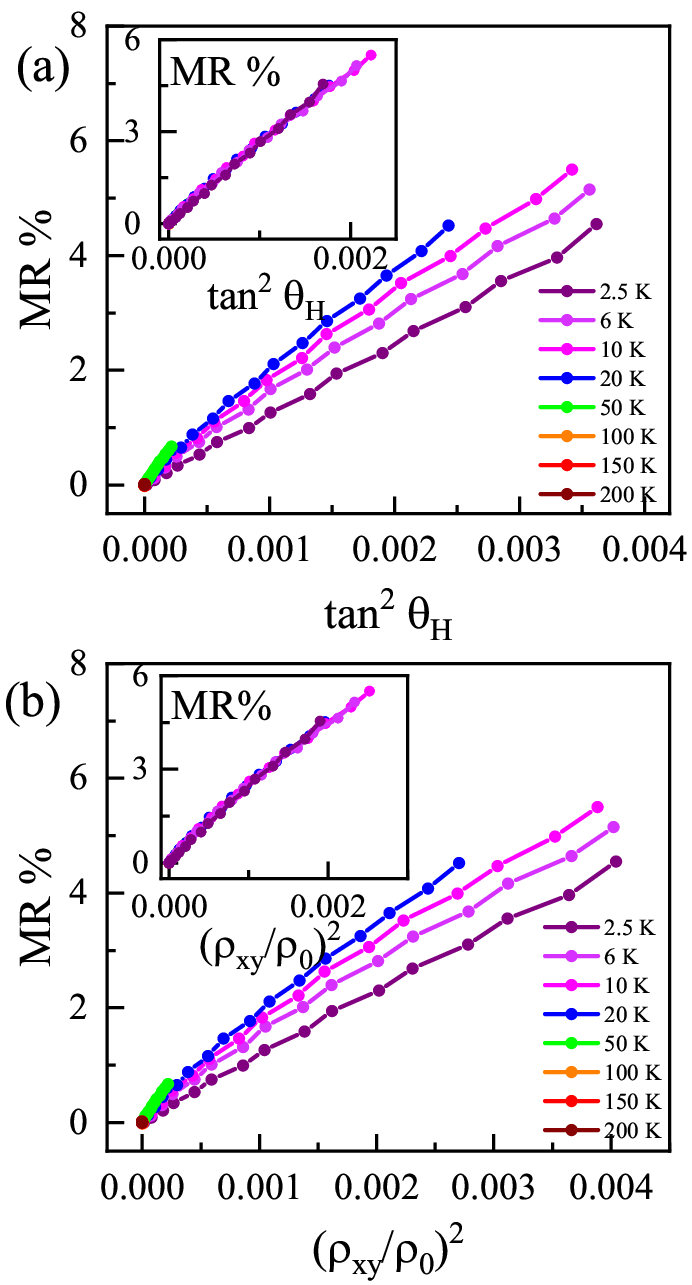}
\caption{Alternative scalings used for the MR in VSe$_2$ sample. (a) MR plotted as a function of square of tangent of Hall angle $tan^{2} \theta_H$ = $(\rho_{xy}/\rho_{xx})^2$. (b) MR plotted as a function of $(\rho_{xy}/\rho_{0})^2$. Corresponding insets show the scaled MR by using the relation MR \% = $\gamma_H tan^{2} \theta_H$ and  MR \% = $\gamma_H (\rho_{xy}/\rho_{0})^2$ respectively.}
\label{fig:Figure3}
\end{figure}

We have plotted MR as a function of tan$^2\theta_H$ and $(\rho_{xy}/\rho_{0})^2$ in figure \ref{fig:Figure3} (a) and (b) respectively. The MR curves at different temperatures scale with both tan$^2\theta_H$ and $(\rho_{xy}/\rho_{0})^2$ above $\textit{T}_{CDW}$, but shows deviation below $\textit{T}_{CDW}$. Corresponding insets show the MR curves scaled in complete \textit{T} range by tuning the $\gamma_H$ values. The violation of (modified) Kohler's rule in VSe$_2$ may be attributed to multi-band effect and the formation of energy gap at Fermi surface in charge density wave state.

\begin{figure} [t]
\includegraphics[width = \columnwidth, height = 12 cm]{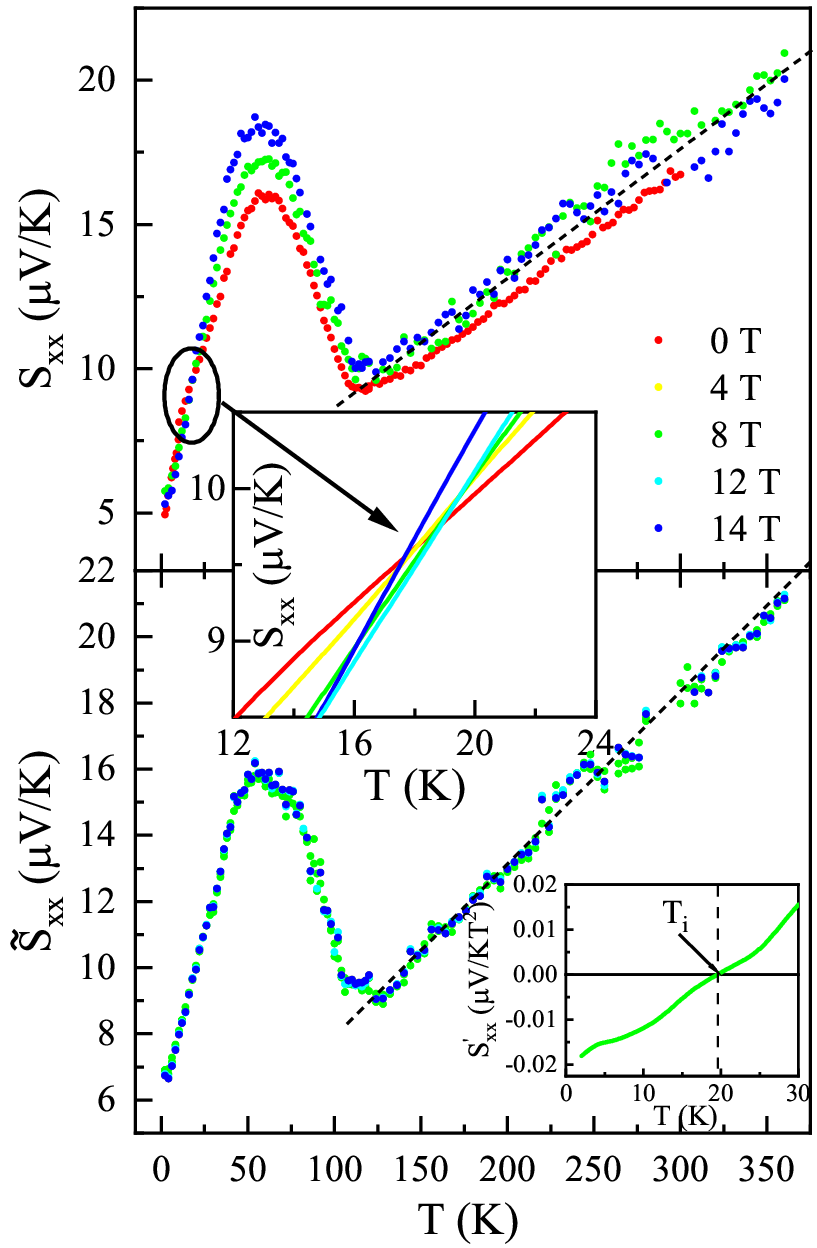}
\caption{Temperature dependence of S$_{xx}$ (upper panel) and $\widetilde{S}_{xx}$ obtained from scaling of isosbestic point (lower panel). The inset shows the crossover at isosbestic point at $\sim{20}$ K. Lower inset shows the first derivative of Seebeck coefficient by squared magnetic field.}
\label{fig:Figure4}
\end{figure} 

The temperature dependence of longitudinal Seebeck coefficient ($\it{S_{xx}}$ = $E_x/|\nabla T_x|$) for \textit{T} = 2 - 350 K at different \textit{H} is shown in the upper panel of figure \ref{fig:Figure4}. The zero field, room temperature ($\it{T}$ = 300 K) $\it{S_{xx}}$ value is $\sim$ 17 $\mu$V/K, which matches with the reported values  \cite{van1978magnetic, yadav2010electronic, zhou2016vanadium}. The $\it{S_{xx}}(T)$ is linear above $\textit{T}_{CDW}$, with the slope (= \textit{dS$_{xx}$/dT}) $\sim$ 0.043 $\mu$V/K$^2$ (highlighted by dashed black line). The Seebeck data show fluctuation at high field near room temperature, that are likely attributable to the measurement noise rather than any significant feature of the material. The linear dependence of $\it{S_{xx}}$ above $\textit{T}_{CDW}$ falls within realm of free electron gas approximation, as also given by Mott relation \cite{behnia2004thermoelectricity},

\begin{equation}\\
S_{xx}/T = \pm \frac{\pi^{2}}{2} \frac{k_B}{e} \frac{1}{T_F} 
\label{eq5}
\end{equation}

where $k_B$ is Boltzmann’s constant, e is electronic charge, and $T_F$ is the Fermi temperature. The $\it{S_{xx}/T}$ value when extrapolated to zero-temperature gives $T_F$ value of 5.5 $\times$ 10$^3$ K and Fermi energy $E_F$ of 0.47 eV for VSe$_2$. The linear \textit{T} dependence for both $\it{S_{xx}}$ and $\rho(T)$ is akin to the linear resistivity observed in the correlated electron system like cuprate superconductor and various heavy fermion systems \cite{PhysRevB.45.5001, anderson1997theory, PhysRevB.56.R4317, naik2007crossover, takagi1992systematic}.

The $\it{S_{xx}}$ exhibits a sharp rise at $\textit{T}_{CDW}$ followed by a broad maximum  centred at $\sim$ 60 K for \textit{H} = 0. This increase in $\it{S_{xx}}$ reflects the partial opening of the gap in the electron-like band, specifically the V 3d$_{z^2}$ band in VSe$_2$. Additionally, the rapid drop of $\it{S_{xx}}$ below 60 K can be attributed to the gap formed in the hole-like band, which corresponds to the Se 4p band in VSe$_2$ \cite{zhou2016vanadium}. The anomaly observed in $\it{S_{xx}}$ is much more pronounced compared to that observed in the \textit{$\rho (T)$} because the $\it{S_{xx}}$ exhibits greater sensitivity to the change in Density of states (DOS) at the Fermi surface. The positive value of $\it{S_{xx}}$ in VSe$_2$ indicates that holes dominate as the charge carriers in this material. The fact that $\it{S_{xx}}$ and $R_H$ have opposite signs suggests the presence of multi-band conduction in VSe$_2$. The density functional theory (DFT) calculations and angle resolved photo emission spectroscopy (ARPES) studies reveal the existence of the hole pockets at the center of the Brillouin zone and a cylinderical electronic Fermi surface at the zone corners in VSe$_2$ \cite{zhou2016vanadium, PhysRevLett.40.1155}. 

Though, we do not observe any appreciable change in the $\textit{T}_{CDW}$ with the application of magnetic field, the maxima at $\sim$ 60 K shift towards lower \textit{T} at higher fields. We believe that the applied magnetic field affects the low temperature scattering, instead of the CDW state. Further, we observed a crossover of $\it{S_{xx}}$ curves taken at different \textit{H}, at $\textit{T}_i$ $\sim$ 20 K (Inset of figure \ref{fig:Figure4}). Remarkably, at this temperature $\it{S_{xx}}$ is independent of magnetic field, showing the existence of isosbestic point for $\it{S_{xx}}$ \cite{cohen1962588, greger2013isosbestic}. It is to mention that isosbestic point (\textit{T$_{i}$}) has been observed in various physical quantities in a variety of compounds \cite{PhysRevB.77.024524, PhysRevB.69.165104, PhysRevLett.39.1098, PhysRevB.81.235130, khoroshilov2016isosbestic, greger2013isosbestic}. The manner in which different curves intersect at an isosbestic point \textit{T$_{i}$}(\textit{H}), determines its type, and there are three main types, (i) Globally exact isosbestic point: This is an ideal case where the curves crossing points \textit{T$_{i}$} (\textit{H}) remain exactly at the same position \textit{T}, regardless of the value of \textit{H}. (ii) Locally exact isosbestic point: In this case, the crossing points of the curves remain fixed only in a small region around a specific value of \textit{H}, denoted as \textit{H$_{0}$}. So, the curves might move if \textit{H} deviates significantly from \textit{H$_{0}$}. This local stability is often depicted by the curve having an extremum or a higher-order stationary point at \textit{H$_{0}$}.  (iii) Approximate isosbestic point: Here, the crossing points of the curves exhibit a weak dependence on \textit{H}. While there might be a general trend of crossing points staying in proximity, their positions are not entirely unaffected by the value of \textit{H} \cite{greger2013isosbestic}. The observed $\it{S_{xx}}$ behavior in VSe$_{2}$ aligns with the characteristics of locally exact isosbestic point. Therefore, it is reasonable to employ the scaling established for locally exact isosbestic point \cite{greger2013isosbestic}. Consequently the $\it{S_{xx}}$ can be approximated as

\begin{equation}
S_{xx}(T,H) = S_{xx}(T,H = 0) + H^2 . S'_{xx} (T) + O (H^4)
\label{eq6}
\end{equation}

where $S'_{xx}(\textit{T})$ = $\frac{S_{xx} (T,H_{1}) - S_{xx} (T,H_{2})}{H_{1}^{2} - H_{2}^{2}}$ is the first order derivative of $\it{S_{xx}}$ by squared magnetic field ($\it{dS_{xx}/dH^{2}}$). Therefore, the scaled Seebeck coefficient in the vicinity of isosbestic point can be expressed as:

\begin{equation}
\widetilde{S}_{xx}(T,H) = S_{xx}(T,H = 0) \approx S_{xx} (T,H) - H^2 . S_{xx}^{'} (T)
\label{eq7}
\end{equation}

The lower panel of figure \ref{fig:Figure4} represents the value of Seebeck coefficient $\widetilde{S}_{xx} (T,H)$, obtained from the scaling for different value of \textit{H} using equation \ref{eq7}. To deduce the value of $S'_{xx} (\textit{T})$, we chose $\it{H_1}$ = 4 Tesla and $\it{H_2}$ = 14 Tesla. Evidently, after scaling all the curves overlap onto a single curve validating the realisation of locally exact isosbestic point in the CDW state of VSe$_2$. The inset in the lower panel of figure \ref{fig:Figure4} depicts the temperature dependence of first order derivative of $\it{S_{xx}}$ with squared magnetic field in the neighbourhood of isosbestic point ($\it{T_{i}}$ = 20 K). The $\it{T_{i}}$ separates the temperature range into positive (\textit{T} $\geq$ 20 K) and negative (\textit{T} $<$ 20 K) $\it{S_{xx}}$ regimes. It is worth mentioning that a weak hump centered around 20 K was also observed in VSe$_2$ single crystal by Yadav \textit{et al.} and Zhou \textit{et al.} \cite{yadav2010electronic, zhou2016vanadium}. The crossover in $\it{S_{xx}}$ at $\it{T_{i}}$ might be attributed to the interplay between \textit{H} and the electronic band structure of the material. The Seebeck coefficient is directly linked to the density of state (DOS) and the energy dependent mobility of the charge carriers within the material. However, the exact mechanisms for the observation of Isobestic point in $\it{S_{xx}}$ may vary depending on the specific material and its electronic properties.

\begin{figure} 
\centering
\includegraphics[width = 7 cm, height = 10 cm]{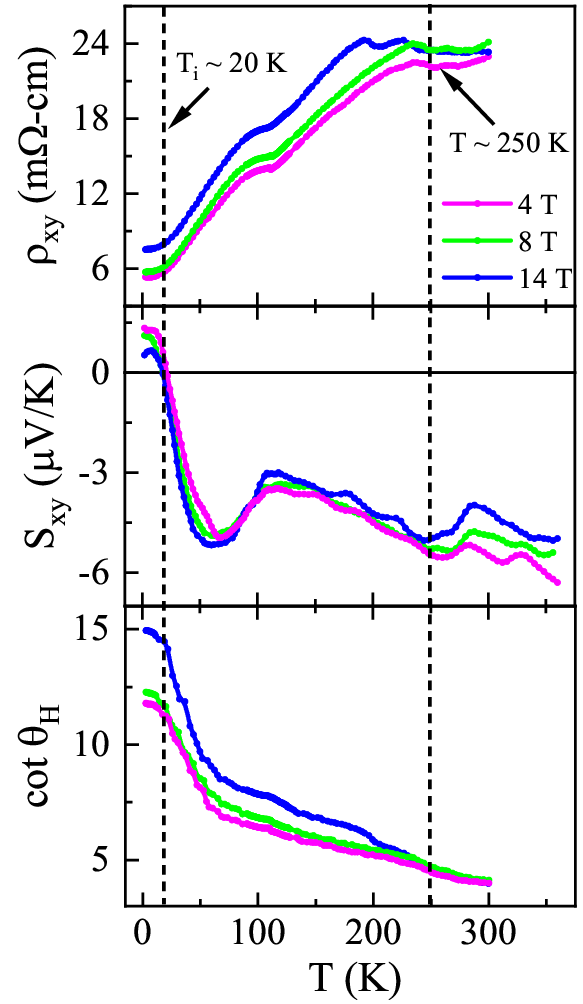}
\caption{Temperature dependence of Hall resistivity $\rho_{xy}$ (panel-I), transverse Seebeck coefficient S$_{xy}$ (panel-II), and Hall angle (panel-III).}
\label{fig:Figure5}
\end{figure}

In figure \ref{fig:Figure5}, we have shown the temperature dependence of Hall resistivity ($\rho_{xy}$), transverse Seebeck coefficient ($\it{S_{xy}}$ = $E_y/|\nabla T_x|$), and Hall angle (cot $\theta_H$ = $\rho_{xx}/\rho_{xy}$) at 4, 8, and 14 Tesla fields. The results reveal two additional features: (i) Anomaly near \textit{T} $\sim$ 250 K: Both $\rho_{xy}$ and $\it{S_{xy}}$ exhibit an anomaly around \textit{T} $\sim$ 250 K for \textit{H} = 4 Tesla, and Hall angle plots at different fields begin to diverge at the same temperature. (ii) Anomaly at $\it{T_{i}}$ = 20 K: All three measurements show anomalies at the isosbestic temperature. The CDW transition is visible in both $\rho_{xy}$ and $\it{S_{xy}}$. Below 250 K, $\it{S_{xy}}$ exhibits linear \textit{T} dependence down to $\textit{T}_{CDW}$. It is remarkable that similar to $\it{dS_{xx}/dH^{2}}$, the $\it{S_{xy}}$ also changes sign at $\it{T_{i}}$ = 20 K. The way Hall angle responds to the variation in temperature in the presence of magnetic field, along with the material's resistivity showing linear \textit{T} dependence, are key signatures for the material's behavior deviating from a Fermi liquid. The Hall angle determines the deviation of the mobile charge carriers' motion from their original trajectory due to the Lorentz force exerted by the magnetic field. Anderson proposed that Hall effect in pure systems can be realised by differentiating between the relaxation rates of carrier motion parallel and perpendicular to the Fermi surface \cite{anderson1991hall}. He gave the expression for the Hall angle as  cot$\theta_H$ = a + b T$^{2}$. We observe that Hall angle cot $\theta_H$ does not show a $\textit{T}^2$ dependence for our VSe$_2$ sample. The temperature dependence of $\rho_{xx}$ and $\it{S_{xx}}$, Hall angle behavior and requirement for the modified Kohler's rule point towards the non-Fermi liquid transport in VSe$_2$.  
\section{Conclusion}

We have demonstrated the influence of magnetic field on the electrical and thermal transport properties of 1T-VSe$_2$. The low temperature ($\textit{T} < \textit{T}_{CDW}$) magnetoresistance deviates from Kohler's rule.  Interestingly, the magnetoresistance value increases with increasing temperature up to \textit{T} $\sim$ 20 K, followed by a decreasing trend on further increase in the temperature. The Kohler slope ($\kappa$) and the relative change in carrier density induced by thermal excitations ($n_T$) also exhibit abrupt changes at \textit{T} = 20 K. Furthermore, the longitudinal Seebeck coefficient curves measured at different magnetic fields, show a locally exact isosbestic point at the same temperature. Anomalies are observed in the Hall resistivity, transverse Seebeck coefficient and the Hall angle near \textit{T} = 20 K. The finding of an isosbestic point in VSe$_2$ presents an exciting opportunity for future theoretical investigation into the band structure particularly near $\it{T_{i}}$ = 20 K to elucidate the underlying mechanism driving these unusual observations.

\section{Acknowledgment}
We acknowledge Advanced Material Research Center (AMRC), IIT Mandi for the experimental facilities. Sonika, SG and PK acknowledge IIT Mandi and MoE, India for the HTRA fellowship. CSY acknowledges SERB-DST (India) for the CRG grant (CRG/2021/002743).

\section{Author contributions}
CSY conceived and supervised the project. Sonika conducted the experiments with the help of SG and PK and A. Taraphder helped us analysing and conceptualizing the results. CSY and Sonika did the complete analysis of data and manuscript writing. 

\section{conflicts of interest}
The authors declare that they have no known competing financial interests or personal relationships that could have appeared to influence the work reported in this research.

 
\bibliography{VSe2}

\end{document}